# Differentiation of functionals with variables coupled by constraints: Analysis through a fluid-dynamical model

Tamás Gál

Department of Theoretical Physics, University of Debrecen, 4010 Debrecen, Hungary
Email address: galt@phys.unideb.hu

**Abstract:** Analysing an application in liquid film dynamics, a guide for obtaining the corresponding constrained functional derivatives for constraints coupling the functional variables is given. The use of constrained derivatives makes the proper account for constraints possible in time-dependent, nonequilibrium physical theories, with physical equations not emerging as Euler-Lagrange equations, which is especially relevant with respect to the dynamics of complex liquids.

In many fields of physics, requirement of the conservation of some extensive property limits the change of physical variables. If functional derivatives are involved in the equations that govern the change of physical variables, an account for the constraints needs to be made in functional differentiation. In the case the physical equation emerges as an Euler equation, from a variational procedure, determining the physical variable(s) as stationary point(s) of some functional $A[\rho]$, the well-known method of Lagrange multipliers gives an appropriate tool for the account for a constraint $C[\rho] = C$, giving

$$\frac{\delta A[\rho]}{\delta \rho(x)} - \mu \frac{\delta C[\rho]}{\delta \rho(x)} = 0$$

for the given physical variable $\rho(x)$, where the multiplier $\mu$ is determined by the equation itself and the constraint. Some more general treatment of constraints, however, is necessary for other cases. For, the introduction of an undetermined "Lagrange" multiplier in itself generally only gives the given physical equation enough freedom to be adjustable to be in accordance with the constraint but does not fix the solution (the physical solution), allowing a range of unphysical solutions.

As an example, consider the equations of motion

$$\frac{\partial h(x)}{\partial t} = \frac{1}{3\eta} \nabla \cdot h^3(x) \nabla \left( \frac{\delta F_T[h,\phi]}{\delta h(x)} - \mu_1 - \mu_2 \phi(x) \right) \quad (1a)$$

and

$$h(x)\frac{\partial \phi(x)}{\partial t} = b^3 \nabla \cdot \left\{ h(x) M(x) \nabla \left( \frac{\delta F_T[h,\phi]}{\delta \phi(x)} - \mu_2 h(x) \right) \right\} + \frac{\phi(x) h^3(x)}{3\eta} \nabla \left( \frac{\delta F_T[h,\phi]}{\delta h(x)} - \mu_1 - \mu_2 \phi(x) \right) \cdot \nabla \phi(x) \quad (1b)$$

for the height $h(x)$ and composition $\phi(x)$ (not denoting time-dependence for simplicity) in the dynamical model of thin liquid films [1] proposed by Clarke [2] for simultaneous dewetting and phase separation in binary mixtures [3], where the two multipliers $\mu_1$ and $\mu_2$ correspond to the constraints

$$\int h(x)\,dx = N_h \tag{2}$$

and

$$\int \phi(x)h(x)\,dx = B \tag{3}$$

of volume and material conservation, respectively, though $\mu_1$ playing no role in the model due to the spatial gradient acting on it. It follows from the form of Eqs.(1) that, irrespective of the $\mu$'s, the solution of them automatically satisfies Eqs.(2) and (3), i.e. the equations

$$\frac{\partial}{\partial t}\int h(x;t)\,dx = 0 \tag{4}$$

and

$$\frac{\partial}{\partial t}\int \phi(x;t)h(x;t)\,dx = 0 \ . \tag{5}$$

The role of the $\mu$'s is to adjust the gradient of $F_T[h,\phi]$, $\left(\left\{\frac{\delta F_T}{\delta h(x)}\right\},\left\{\frac{\delta F_T}{\delta \phi(x)}\right\}\right)$, to be in accordance with the constraints (2) and (3) (only with the proper $\mu[C[\rho];F[\rho]]$'s the solution $h(x;t)$ and $\phi(x;t)$ of Eqs.(1) will be physical). For, the behaviour of $F_T[h,\phi]$ only over the domain determined by Eqs.(3) and (4) may govern the motion of $h(x)$ and $\phi(x)$, implying that the equations of motion for $h(x)$ and $\phi(x)$ have to be invariant under the replacement of $F_T[h,\phi]$ with a different functional $F'_T[h,\phi]$ that equals $F_T[h,\phi]$ over the domain Eqs.(3)-(4) (which leads to the K-equality condition on the derivative of $F_T[h,\phi]$; see later). Note that $\mu_2$, of course, is expressable from Eqs.(1a) and (1b) as e.g.

$$\mu_2 = \frac{1}{3\eta\int h(x)\{\nabla\cdot h^3(x)\nabla\phi(x)\}dx}\int h(x)\left\{\nabla\cdot h^3(x)\nabla\frac{\delta F_T[h,\phi]}{\delta h(x)} - \frac{\partial h(x)}{\partial t}\right\}dx \ , \tag{6}$$

but the proper expression for $\mu_2$ cannot be expressed from Eqs.(1) in that way, contrary to the case a physical equation emerges as an Euler equation. The necessity of a modification of $\dfrac{\delta F_T[h,\phi]}{\delta h(x)}$ and $\dfrac{\delta F_T[h,\phi]}{\delta \phi(x)}$ under constraints is directly shown by the case of equilibrium, where $h(x;t)$ and $\phi(x;t)$ are constant in time; since, without a modification, Eqs.(1) cannot give the Euler-Lagrange equations

$$\dfrac{\delta F_T[h,\phi]}{\delta h(x)} = \mu_1 + \mu_2 \phi(x) \tag{7a}$$

and

$$\dfrac{\delta F_T[h,\phi]}{\delta \phi(x)} = \mu_2 h(x) \tag{7b}$$

of equilibrium, stemming from the minimization of the free energy $F_T[h,\phi]$ under the constraints (2) and (3).

Recently, the proper expression for the multiplier $\mu$ in (a general)

$$\dfrac{\delta A[\rho]}{\delta \rho(x)} - \mu \dfrac{\delta C[\rho]}{\delta \rho(x)} \tag{8}$$

to account for constraints of the form

$$\int f(\rho(x))dx = K \tag{9}$$

(with invertible $f$, which may have an explicit $x$-dependence as well), where $\dfrac{\delta C[\rho]}{\delta \rho(x)} = f^{(1)}(\rho(x))$, has been derived [4], obtaining

$$\dfrac{\delta A[\rho]}{\delta_K \rho(x)} = \dfrac{\delta A[\rho]}{\delta \rho(x)} - \left(\dfrac{1}{K}\int \dfrac{f(\rho(x'))}{f^{(1)}(\rho(x'))}\dfrac{\delta A[\rho]}{\delta \rho(x')}dx'\right)f^{(1)}(\rho(x)) \tag{10}$$

(introducing the concept of constrained derivatives), which gives

$$\dfrac{\delta A[\rho]}{\delta_L \rho(x)} = \dfrac{\delta A[\rho]}{\delta \rho(x)} - \left(\dfrac{1}{L}\int \rho(x')\dfrac{\delta A[\rho]}{\delta \rho(x')}dx'\right)g(x) \tag{11}$$

for linear $C[\rho]$'s, $\int g(x)\rho(x)dx = L$. K-conserving differentiation has been extended to treat multiple (simultaneous) K-constraints (9) as well [5]. Clarke [2] applied the method of K-conserving differentiation in his above model to determine proper $\mu$'s to account for the constraints (2) and (3), and verified the obtained $\mu$'s theoretically as well with the help of an earlier result in his model [6]. The application of K-conserving differentiation in the case of constraints coupling variables of the differentiated functional, like Eq.(3), however, raises some questions, which will be answered in the following, giving a general guide for obtaining the constrained derivatives needed.

Clarke had obtained the proper expressions

$$\mu_1 = \frac{1}{N_h}\left\{\int h(x')\frac{\delta F_T[h,\phi]}{\delta h(x')}dx' - \int \phi(x')\frac{\delta F_T[h,\phi]}{\delta \phi(x')}dx'\right\} \tag{12a}$$

and

$$\mu_2 = \frac{1}{B}\int \phi(x')\frac{\delta F_T[h,\phi]}{\delta \phi(x')}dx' , \tag{12b}$$

i.e. the derivatives

$$\frac{\delta F_T[h,\phi]}{\delta_K h(x)} = \frac{\delta F_T[h,\phi]}{\delta h(x)} - \frac{1}{N_h}\left\{\int h(x')\frac{\delta F_T[h,\phi]}{\delta h(x')}dx' - \int \phi(x')\frac{\delta F_T[h,\phi]}{\delta \phi(x')}dx'\right\} - \frac{\phi(x)}{B}\int \phi(x')\frac{\delta F_T[h,\phi]}{\delta \phi(x')}dx' \tag{13a}$$

and

$$\frac{\delta F_T[h,\phi]}{\delta_K \phi(x)} = \frac{\delta F_T[h,\phi]}{\delta \phi(x)} - \frac{h(x)}{B}\int \phi(x')\frac{\delta F_T[h,\phi]}{\delta \phi(x')}dx' \tag{13b}$$

($K$ denoting the constraint (2)-(3) here), for his model via finding the proper decomposition of the functional variables [7],

$$h(x) = \tilde{h}(x)\frac{N_h}{\int \tilde{h}(x')dx'} \tag{14a}$$

and

$$\phi(x) = \tilde{\phi}(x)\frac{B}{\int \tilde{\phi}(x')\tilde{h}(x')\frac{N_h}{\int \tilde{h}(x'')dx''}dx'} , \tag{14b}$$

using the conditions [4] that (i) for a $\rho_K(x)$ (satisfying the given constraint), the decomposition $\rho[\tilde{\rho}, K]$ should give back $\rho_K(x)$, and (ii) $\rho[\tilde{\rho}, K](x)$ should satisfy the constraint for any $\tilde{\rho}(x)$. The decomposition (14) yields the K-constrained derivatives (13) by [4]

$$\frac{\delta F_T[\rho]}{\delta_K \rho(x)} = \left( \frac{\delta F_T[\rho[\tilde{\rho}, K]]}{\delta \tilde{\rho}(x)} \right)_K \bigg|_{\tilde{\rho}=\rho} . \qquad (15)$$

That procedure can alternatively be viewed [8] as obtaining the constrained derivatives through

$$\frac{\delta F_T[\rho]}{\delta_K \rho(x)} = \frac{\delta F_T[\rho_K[\rho]]}{\delta \rho(x)} , \qquad (16)$$

where $\rho_K[\rho] = (h_K[h,\phi], \phi_K[h,\phi])$ is the extension

$$h_K[h,\phi](x) = h(x) \frac{N_h}{\int h(x')dx'} \qquad (17a)$$

and

$$\phi_K[h,\phi](x) = \phi(x) \frac{B}{\int \phi(x') h(x') \frac{N_h}{\int h(x'')dx''} dx'} \qquad (17b)$$

of the functional variables from the *K*-restricted domain, satisfying conditions corresponding to the above conditions (i) and (ii). The decomposition, or extension, of $h(x)$ and $\phi(x)$ yielding (13) may be considered as applying the constraint (4) to $\phi(x)$ while letting $h(x)$ vary free of it (but under (2)). The question immediatelly arises then as whether the extension/decomposition corresponding, e.g., to applying (3) also to $h(x)$, beside (2), is a proper extension/decomposition. On the basis of the two conditions (i) and (ii), the following extension emerges for that case with the help of the extension [5] obtained for simultaneous linear K-constraints:

$$h_K[h,\phi](x) = h(x)\frac{N_h}{\int h(x')dx'} - \frac{\sigma(x)}{\int \phi(x')\sigma(x')dx'}\left(\frac{N_h}{\int h(x')dx'}\int \phi(x')h(x')dx' - B\right) \quad (18a)$$

and

$$\phi_K[h,\phi](x) = \phi(x) \quad (18b)$$

(with $\sigma(x)$ an arbitrary function that integrates to zero), giving

$$\frac{\delta F_T[h,\phi]}{\delta_K h(x)} = \frac{\delta F_T[h,\phi]}{\delta h(x)} - \frac{1}{N_h}\left\{\int h(x')\frac{\delta F_T[h,\phi]}{\delta h(x')}dx' - \frac{B}{\int \phi\sigma}\int \sigma(x')\frac{\delta F_T[h,\phi]}{\delta h(x')}dx'\right\} - \frac{\phi(x)}{\int \phi\sigma}\int \sigma(x')\frac{\delta F_T[h,\phi]}{\delta h(x')}dx' \quad (19a)$$

and

$$\frac{\delta F_T[h,\phi]}{\delta_K \phi(x)} = \frac{\delta F_T[h,\phi]}{\delta \phi(x)} - \frac{h(x)}{\int \phi\sigma}\int \sigma(x')\frac{\delta F_T[h,\phi]}{\delta h(x')}dx' . \quad (19b)$$

Other formulae arise as well if the constraint (3) is divided in some proportion between $h$ and $\phi$ (or, as for simultaneous constraints on a single variable, with the transformation in Eq.(18a) $h(x) \to \phi(x)h(x)$, $\phi(x) \to 1$ in the integrands, and $\sigma(x) \to \sigma(x)/\phi(x)$ and $N_h \leftrightarrow B$).

Ambiguity emerging from the above conditions (i)-(ii) for a proper extension $\rho_K[\rho]$ appears even in the case of one functional variable with a single K-constraint, and a third condition, degree-zero K-homogeneity (that is, for linear K-constraints, normal degree-zero homogeneity), is what makes $\rho_K[\rho]$, that gives the proper $\frac{\delta}{\delta_K \rho}$, unique [8,5]. (For single K-constraints, conditions (i) and (iii) alone also yield the proper, unique $\rho_K[\rho]$ [8].) Conditions (i) and (ii) in themselves lead to an (ambiguous) derivative that fulfils only the K-equality condition (namely, the condition that two functionals that are equal over a *K*-restricted domain should have equal *K*-conserving derivatives over that domain), and condition (iii) is needed to fulfil the other condition for $\frac{\delta}{\delta_K \rho}$: the K-independence condition, namely, that $\frac{\delta}{\delta_K \rho}$ has to yield $\frac{\delta}{\delta \rho}$ for *K*-independent functionals. The question is as how condition (iii) applies for the present case of a constraint coupling two variables of the same functional. First, condition (iii) applies for the extension of the product of the two variables $h(x)$ and $\phi(x)$, yielding the unique

$$(\phi_K[h,\phi] h_K[h,\phi])(x) = \phi(x) h(x) \frac{B}{\int \phi(x') h(x') dx'} , \qquad (20)$$

since the $\frac{\delta}{\delta_K \rho}$-formula has to be valid also for $G[\phi h]$, with *G* being a one-variable functional. Second, condition (iii) applies for $\phi_K[h_K, \phi]$, since it yields a $F_T[h_K, \phi_K[h_K, \phi]]$ that is independent of *M* in its variable $\phi(x)$, i.e. invariant under changes $\lambda \phi(x)$ of $\phi(x)$ (at a fixed $h_K(x)$), which leave $\frac{\phi(x)}{\int \phi(x') h_K(x') dx'}$ of $\phi(x)$ unchanged. (For a fixed $h_K(x)$, the two-variable case can be considered as a single-variable case, with one linear constraint, Eq.(3).) Thus,

$$\phi_K[h_K, \phi](x) = \phi(x) \frac{B}{\int \phi(x') h_K(x') dx'} . \qquad (21)$$

From Eqs.(20) and (21) then it follows that among the possible extensions of $h_K(x)$ that satisfy conditions (i) and (ii), Eq.(17a) is the proper one, that is, Eqs.(17) is the full proper extension (i.e. Eqs.(14) is the proper decomposition), yielding the *K*-constrained derivatives (13). Note that, of course, condition (iii) applies for $h_K[h, \phi_K]$ (with $\phi_K(x)$ fixed) as well, but does not yield a unique $h(x)$-dependence, because of the simultaneous constraints on $h(x)$.

Having the constrained derivative formulae for the constraint Eqs.(2) and (3), the question naturally arises as how the constrained derivative looks like for a constraint (3) alone. In that case,

$$h_K[h,\phi](x) = h(x)\left(\frac{B}{\int \phi(x')h(x')\,dx'}\right)^n \quad (22a)$$

and

$$\phi_K[h,\phi](x) = \phi(x)\left(\frac{B}{\int \phi(x')h(x')\,dx'}\right)^{1-n}, \quad (22b)$$

with any $n$, fulfil Eq.(20) (and satisfy conditions (i) and (ii)), yielding

$$\frac{\delta F[h,\phi]}{\delta_K h(x)} = \frac{\delta F[h,\phi]}{\delta h(x)} - \frac{\phi(x)}{B}\left\{n\int h(x')\frac{\delta F[h,\phi]}{\delta h(x')}\,dx' + (1-n)\int \phi(x')\frac{\delta F[h,\phi]}{\delta \phi(x')}\,dx'\right\} \quad (23a)$$

and

$$\frac{\delta F[h,\phi]}{\delta_K \phi(x)} = \frac{\delta F[h,\phi]}{\delta \phi(x)} - \frac{h(x)}{B}\left\{n\int h(x')\frac{\delta F[h,\phi]}{\delta h(x')}\,dx' + (1-n)\int \phi(x')\frac{\delta F[h,\phi]}{\delta \phi(x')}\,dx'\right\}; \quad (23b)$$

however, $h_K[h,\phi_K]$ and $\phi_K[h_K,\phi]$ both cannot fulfil condition (iii), that is, they cannot be homogeneous of degree zero simultaneously. That means that no $K$-constrained derivative exists for the case of Eq.(3), which shows the importance of a second constraint on at least one of the functional variables.

In the case of a normalization-conservation constraint on the second variable as well,

$$\int \phi(x)\,dx = N_\phi, \quad (24)$$

the treatment [5] of simultaneous constraints on one variable is needed. For that, it is important that in the extension (20), an additional term $+\sigma'(x)\xi\left(\int \phi(x')h(x')\,dx' - B\right)$ is also allowed, where $\sigma'(x)$ is an arbitrary function that integrates to zero and is homogeneous of degree zero in $\phi(x)h(x)$, and $\xi$ is an arbitrary function for which $\xi(0) = 0$ [5]. (In the case of the constraint (2)-(3), that term has no role, as for single constraints either.) Then the proper extension is

$$h_K[h,\phi](x) = h(x)\frac{N_h}{\int h(x')\,dx'} \qquad (25a)$$

and

$$\phi_K[h,\phi](x) = \phi(x)\frac{B}{\int \phi(x')h(x')\frac{N_h}{\int h(x'')\,dx''}\,dx'} - \frac{\sigma(x)/h(x)}{\int(\sigma(x')/h(x'))\,dx'}\left(\frac{B}{\int \phi(x')h(x')\frac{N_h}{\int h(x'')\,dx''}\,dx'}\int \phi(x')\,dx' - N_\phi\right),$$

(25b)

satisfying the three conditions, and giving

$$\frac{\delta F[h,\phi]}{\delta_K h(x)} = \frac{\delta F[h,\phi]}{\delta h(x)} - \frac{1}{N_h}\left\{\int h(x')\frac{\delta F[h,\phi]}{\delta h(x')}\,dx' - \int \phi(x')\frac{\delta F[h,\phi]}{\delta \phi(x')}\,dx' + \frac{N_\phi}{\int\frac{\sigma}{h}}\int\frac{\sigma(x')}{h(x')}\frac{\delta F[h,\phi]}{\delta \phi(x')}\,dx'\right\}$$

$$-\frac{\phi(x)}{B}\left\{\int \phi(x')\frac{\delta F[h,\phi]}{\delta \phi(x')}\,dx' - \frac{N_\phi}{\int\frac{\sigma}{h}}\int\frac{\sigma(x')}{h(x')}\frac{\delta F[h,\phi]}{\delta \phi(x')}\,dx'\right\} \quad (26a)$$

and

$$\frac{\delta F[h,\phi]}{\delta_K \phi(x)} = \frac{\delta F[h,\phi]}{\delta \phi(x)} - \frac{1}{\int\frac{\sigma}{h}}\int\frac{\sigma(x')}{h(x')}\frac{\delta F[h,\phi]}{\delta \phi(x')}\,dx' - \frac{h(x)}{B}\left\{\int \phi(x')\frac{\delta F[h,\phi]}{\delta \phi(x')}\,dx' - \frac{N_\phi}{\int\frac{\sigma}{h}}\int\frac{\sigma(x')}{h(x')}\frac{\delta F[h,\phi]}{\delta \phi(x')}\,dx'\right\}.$$

(26b)

Note that the two variables can be interchanged in Eq.(26), because of the symmetry of the constraints in them, giving a further ambiguity, similarly to the case of two simultaneous constraints on a single functional variable [5].

As a further example, the constraints (2), (3), and

$$\int \chi(x)\phi(x)h(x)\,dx = T, \qquad (27)$$

e.g., on three variables ($h(x)$, $\phi(x)$, and $\chi(x)$), can be mentioned. Then the extension (homogeneous of degree zero in all the three variables) yielding the corresponding constrained functional derivatives is

$$h_K[h,\phi,\chi](x) = h(x)\frac{N_h}{\int h(x')dx'} \;, \tag{28a}$$

$$\phi_K[h,\phi,\chi](x) = \phi(x)\frac{B}{\int \phi(x')h(x')\frac{N_h}{\int h(x'')dx''}dx'} \;, \tag{28b}$$

and

$$\chi_K[h,\phi,\chi](x) = \chi(x)\frac{T}{\int \chi(x')\phi(x')h(x')\frac{B}{\int \phi(x'')h(x'')dx''}dx'} \;. \tag{28c}$$

Finally, the simple coupling

$$\int h(x)\,dx + \int \phi(x)\,dx = N \tag{29}$$

of two functional variables is worth of consideration. In that case, the extension

$$h_K[h,\phi](x) = h(x)\frac{N}{\int h(x')\,dx' + \int \phi(x')\,dx'} \tag{30a}$$

and

$$\phi_K[h,\phi](x) = \phi(x)\frac{N}{\int h(x')\,dx' + \int \phi(x')\,dx'} \tag{30b}$$

gives the proper

$$\frac{\delta F[h,\phi]}{\delta_K h(x)} = \frac{\delta F[h,\phi]}{\delta h(x)} - \frac{1}{N}\left(\int h(x')\frac{\delta F[h,\phi]}{\delta h(x')}dx' + \int \phi(x')\frac{\delta F[h,\phi]}{\delta \phi(x')}dx'\right) \tag{31a}$$

and

$$\frac{\delta F[h,\phi]}{\delta_K \phi(x)} = \frac{\delta F[h,\phi]}{\delta \phi(x)} - \frac{1}{N}\left(\int h(x')\frac{\delta F[h,\phi]}{\delta h(x')}dx' + \int \phi(x')\frac{\delta F[h,\phi]}{\delta \phi(x')}dx'\right) \;. \tag{31b}$$

Here, condition (iii) applies for $(h,\phi)$, that is, only a simultaneous change $(\lambda h, \lambda \phi)$ has to be cancelled in Eq.(30). It is worth pointing out that the formula (31) can be obtained also directly from the formula $\frac{\delta A[\rho]}{\delta_N \rho(x)} = \frac{\delta A[\rho]}{\delta \rho(x)} - \frac{1}{N}\int \rho(x')\frac{\delta A[\rho]}{\delta \rho(x')}dx'$ for the constraint $\int \rho(x)dx = N$, being valid for the discrete case as well [5], and Eq.(29) being a mixture of the continuous and the discrete case of normalization conservation.

In summary, a guide for obtaining the corresponding constrained functional derivatives for constraints coupling the functional variables, bearing particular relevance with respect to the dynamics of complex liquids, has been given, analysing a fluid-dynamical application.

**Acknowledgments:** Grant D048675 from OTKA is gratefully acknowledged.